\documentclass[epj]{svjour}
\usepackage{graphics}
\usepackage{epsfig,amsmath,amsfonts,amssymb,setspace,multirow,textcomp}
\begin{document}
\title{Advanced morphology of VIPERS galaxies}
\subtitle{Gini, M20 and CAS statistics\\ with detailed analysis of Sersic index}
\author{Tugay A.\inst{1}, Gugnin O.\inst{1}, Pulatova N.\inst{2}, Zadorozhna L.\inst{1}
}                     
\offprints{}          
\institute{Taras Shevchenko National Univercity of Kyiv \and Main Astronomical Observatory of National Academy of Science of Ukraine}
\date{Received: 2021 / Revised version: 2022}
\abstract{ 
We calculated morphological parameters for 70821 galaxy from VIPERS survey. These parameters includes Gini, M20, Concentration, Asymmetry and Smoothness. Results correlate with the distribution of these parameters for other simulated and observed samples. We also studied dependence of these parameters with Sersic power index of radial distribution of surface brightness of galaxy image. 
} 
\maketitle

\tableofcontents

\newpage

\section{Introduction}
\label{intro}
\label{sec:1}

VIPERS is major galaxy survey for LSS study \cite{Scodeggio2018}. It contains 94.941 galaxies with $0.5<z<1.0$ from 24 deg$^2$. Their positions are important cosmology information that was used for recovering 3D filament structure in observed volume \cite{Malavasi2017}. But in addition to positions it is important also to study the images of galaxies. In the first approximation galaxy images can be considered as ellipses with radial distribution of surface brightness given by Sersic profile \cite{Krywult2017}. There are more detailed descriptions of images beyond Sersic profile. One of the best sets of advanced morphology parameters are Gini and M20 statistics, Concentration, Asymmetry and Smoothness. These parameters can be calculated by statmorph program that was written by Rodriguez-Gomez on Python in \cite{Rodriguez-Gomez2019} . Our task was to use statmorph code explained there to calculate mentioned parameters for VIPERS galaxies. It is very important task because these parameters can be used for two cosmological studies. The first is the study of galaxy merging history during of evolution of Universe \cite{deRavel2009}. The second is the analysis of influence of environment to galaxy morphology \cite{Kampczyk2013} \cite{Tasca2009}. That means that enviroment should have influence to galaxy formation. We can analyse the result of galaxy formation in the distribution of galaxy parameters. 
Calculation of morphological parametres of galaxies is very important for studying the extragalactical Universe, because, as it was mentioned above, the process of galaxies merging is highly bounded with its morphological features. For these galaxies, their appearence and any physical parametres are highly depended on the morphological types, masses, redshifts, environments, and the previous star formation and merging histories of the individual galaxies\cite{Kampczyk2013}.
It is very useful to study the influence of enviroment on galaxy formation and merging too, because, as it is said in \cite{Tasca2009} there is a corellation between the enviroment and galaxy type, which can be obtained from knowledge of morphological parametres, e.g. early–type galaxies are preferentially found in denser regions than late–type ones\cite{Oemler1974}. Enviromental characteristics can be used in studying not only formation and evolution of galaxies, but also their merging and interactions.

\newpage

\section{Sample}
\label{sec:2}

VIPERS is spectroscopic galaxy survey performed on VIMOS spectroscope at VLT \cite{Scodeggio2018}. Thus redshifts of all studied galaxies were found in this survey. VIPERS consists of W1 and W4 regions of CFHTLS where there are images of galaxies. 

To test Gini and M20 distribution with statmorph we selected 
4659 galaxies from one square degree of W4. This sample corresponds to one single plate from CFHTLS. When we will finish this preliminary analysis we will calculate morphology parameters for all VIPERS galaxies. We think that distribution of morphology parameters for the whole VIPERS sample will be the same. 

Since we took test sample from W4 field and later calculate Sersic index for it, we present here distribution of basic parameters of the whole W4 VIPERS sample.
This sample contains 32.937 galaxies. The values of Sersic index for it is analysed in Section 4 of current work. In the sample there are a number of flags in addition to the values of different parameters. Redshift distribution of W4 galaxies (from right ascension) is presented at Figs 1-2.
Full range of redshift flag is -100..230. Except of small group with flag from -11 to -12, most flags are between 0 and 1: 0.2, 0.4 and 0.5. $z<2.2$, it is upper limit.

\begin{figure*}[h]
\centering
\includegraphics[scale = 0.49]{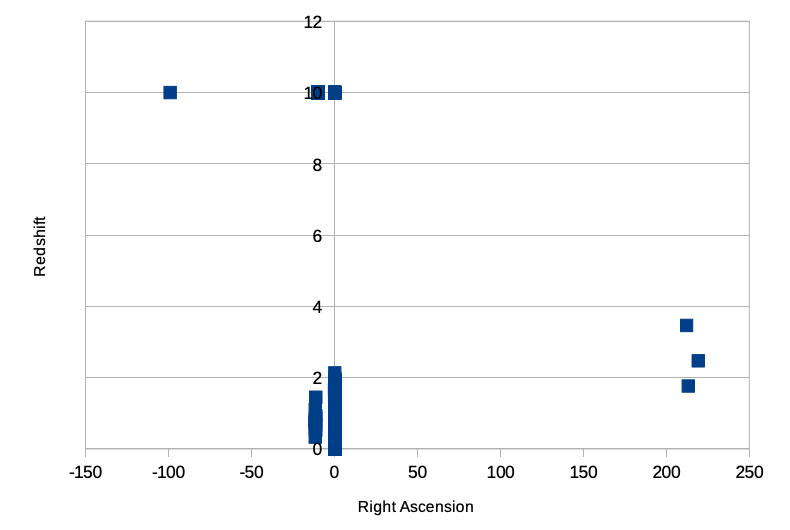}
\caption{RA-z distribution for W4 with flags}
\label{fig:w41}
\end{figure*}

\begin{figure*}[h]
\centering
\includegraphics[scale = 0.49]{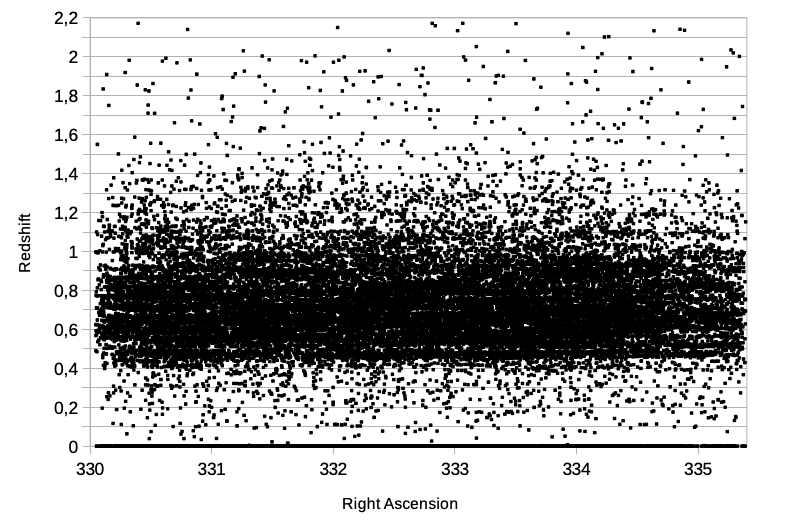}
\caption{RA-z distribution for main part of W4}
\label{fig:w42}
\end{figure*}

\newpage

\begin{figure*}[h]
\centering
\includegraphics[scale = 0.49]{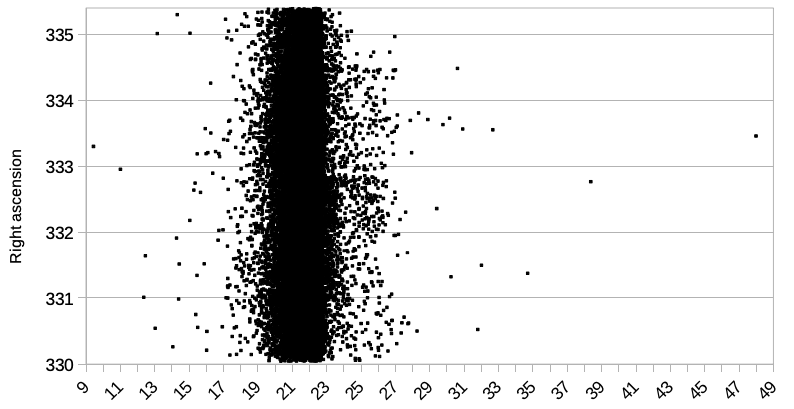}
\caption{Distribution of RA from magnitude}
\label{fig:w43}
\end{figure*}

Elements of LSS in W4 can be seen at Fig. 4.

\newpage

\begin{figure*}[h]
\centering
\includegraphics[scale = 0.49]{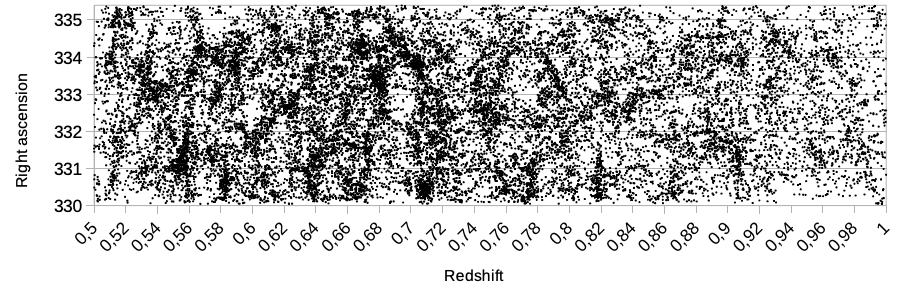}
\caption{LSS for W4 subsample of VIPERS.}
\label{fig:w44}
\end{figure*}

\newpage

\section{Method}
\label{sec:3}

We calculated five morphological parameters for VIPERS galaxies: Gini, M20, Concentration, Asymmetry, Smoothness. 

G parameter is calculated as $$G = \frac{1}{|\overline{X}|n(n-1)}\sum_{i=1}^n{(2i-n-1)|X_i|},$$
where $X_i$ are the flux values of n pixels(\cite{Lotz2004}).
$M_{20}$ is obtained as(\cite{Lotz2004})
$$
M_{20} \equiv \log_{10}\left(\frac{\sum_i\mu_i}{\mu_{tot}}\right) , while \sum_iI_i<0.2I_{tot},
$$
Where $\mu_{tot} = \sum_{i = 1}^n\mu_i = \sum_{i = 1}^nI_i[(x_i-x_c)^2-(y_i-y_c)^2]$. $I_i$ - pixel flux values, ($x_c,y_c$) - galaxy's centre.

Concentration index is calculated(\cite{Bershady2000}):
$$
C = 5\lg\left(\frac{r_{80}}{r_{20}}\right),
$$
where where r20 and r80 are the radii of circular apertures containing 20
and 80 per cent of the galaxy’s light. In this case total flux can be measured in 1.5 Petrosian radii.

The asymmetry index can be calculated by substracting the galaxy image, wich was rotated by $180^\circ$ from the original image\cite{Conselice2000}:
$$
A = \frac{\sum_{i,j}|I_{i,j} - I_{i,j}^{180}|}{\sum_{i,j}|I_{i,j}|} - A_{bgr},
$$
where $A_{bgr}$ is the average background asymmetry.

The smoothness index can be obtained as\cite{Conselice2003}:
$$
S = \frac{\sum_{i,j}|I_{i,j} - I_{i,j}^{S}|}{\sum_{i,j}|I_{i,j}|} - S_{bgr},
$$
where $S_{bgr}$ is the average background smoothness.

To calculate these parameters we used statmorph code  
\cite{Rodriguez-Gomez2019}.
Method of statmorph application was developed basing on statmorph tutorial \footnote{https://nbviewer.jupyter.org/github/vrodgom/statmorph/blob/master/notebooks/tutorial.ipynb}.

	The method has the following stages.
	
1. We selected poststamps from CFHTLS image for all galaxies from test sample with corresponding coordinates. We got poststamps of 100x100 pixels for each VIPERS galaxy in i-band. Generation of test initial image by Sersic model is presented at Fig.\ref{ap1}. According to definitions of morphological parameters \cite{Rodriguez-Gomez2019}, four of them (Gini, M20, Concentration and Smoothness) does not require deviations of image from round shape. Thus for the estimation of errors of Gini, M20, C and S we simulated images with Sersic profile and zero ellipticity. For the estimation of asymmetry errors the superposition of two elliptical Sersic galaxies was used (see chapter 5.2). Distance between the centers of components is 10 pixels. Amplitude (brightness) of second galaxy is two times less than the first.

2. To analyse images with statmorph we approximated point-spread function (PSF) by two-dimensional Gaussian function with $\sigma $ = 2 pixels. Separate analysis of PSF is presented in the Chapter 4.2.

Until formation mock images for tests of errors (Chapters 4, 5.2 and 5.3) we generated separate images of PSF (Fig.\ref{ap1}). PSF is the same for all simulations.
Then the galaxy image was convoluted with PSF (Fig.\ref{ap2})
and random noise was added (Fig.\ref{ap2}).

3. Finding a segmentation map for the image (Fig.\ref{ap3}). 
Next, in order for the program to understand which points in the image are the source and which are not, a segmentation map was built using photutils, an astropy package for astrometry. Photutils provides two functions designed specifically to detect point-like (stellar) sources in an astronomical image. 

4. Smoothing the segmentation map (Fig.\ref{ap3}). We used \textbf{scipy.ndimage.uniform\_filter} function to smooth the shape of segmentation map and reject single-pixel regions from it which are disconnected with the region of the main source. The input of this function is a segmentation map, and the second argument in it is "size", the size of the uniform filter for each of the axes, here is the same 10 for both. 

5. After that we launch statmorph by the command \textbf{statmorph.source\_morphology (image, segmap, gain = 100.0, psf = psf)}, where the first two and fourth arguments were mentioned above. Gain parameter is used for calculation of pixel weights within segmantation map and indicates supposed averane number of counts per pixel inside effective radius.

This is enough to correctly calculate all morphological parameters except the Sersic index, for which the special algorithm was used. This algirithm is described in Chapter 4.

\newpage

\section{Sersic index analysis}

Statmorph uses simple routine for Sersic fitting which can not get parameters of fit in 40\% cases. In these cases statmorh outputs Sersic index n=1 or crashes. So we performed modifications of segmentation map, background and images for these cases to force statmorph output any values for Sersic fit without crash. We stress here once more, that there were no problems with statmorph in caclulating advanced morphological parameters: Gini, M20, Concentration, Asymmetry and Smoothness. These parameters are calculated in very separate procedure based on decomposition of image by power-law momentums. Modifications described below were performed only in our attempt to compare statmorph Sersic fit with the same fit from other works, e.g. GALFIT \cite{Krywult2017}. 

When Sersic fit was failed, we used at first a "fake segmentation map", which is a "mask" with a circle of radius of 15, and an array "round" for additional fake image, with a circle of the same radius, with values of Gaussian image $exp(-(r2/50)^2)$, where r2 is the radius of filling of the array. With the help of this map, cases in which \textbf{np.argmax(segm.areas)} equals 0 was avoided. This occurs because the largest source is being searched for, and sometimes program has issues with this.
Our algorythm of forcing statmorph Sersic fit was the following. 
At first we processed image by statmorph without corrections.
Then there was a check for the Sersic index. 
If it equals 1 or 2, then "fake segmentation map" was used, and
Sersic index was recalculated.
If the entire Sersic fit was failed, then the image was modified by adding Gaussian round image described above. After that, the fit was recalculated again. If it is equals 1 or 2 again, the cycle repeats with the increasing of brightness of Gaussian component, if not, the latest value of Sersic index is saved. So at the output one have a set of morphological parameters calculated using the statmorph program for real galaxies with the best available Sersic fit.

\newpage

\subsection{Bimodality test}

We performed two tests of morphology parameters for our sample. The aim of the first of them was to find a clear division of VIPERS galaxies to elliptical and spiral. This is important to test the method of Sersic index (ns) calculation in statmorph program. To find such bimodality we use B-V color index from VIPERS database. Spiral galaxies must have B-V$<$1.3 and ns$>$0.7. Elliptical galaxies must have B-V$>$1.3 and ns$<$0.7. We used sample of 4388 VIPERS galaxies from one W4 field 60'x60' centered at RA=22h13m18s, DEC=+01d19m00s. Distribution of color and Sersic index for this sample is shown at (Fig. 5). To test bimodality the sample was divided to 11 subsamples (Fig. 6).

\begin{figure*}[h]
\includegraphics[width=1\linewidth]{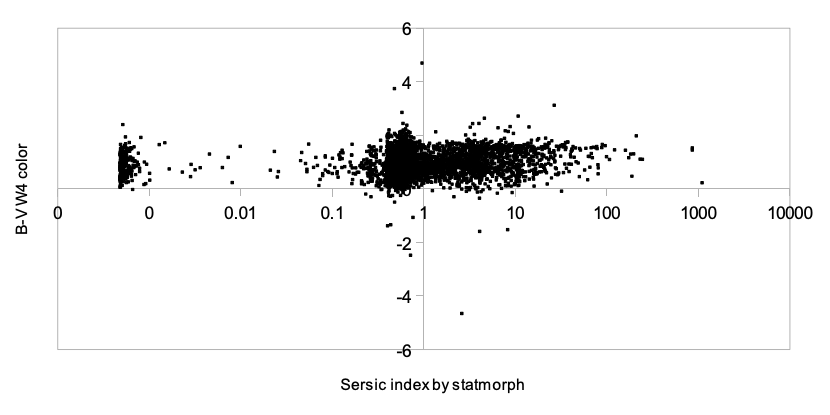}
\caption{Distribution of color and Sersic index for the entire sample.}
\label{fig:bm1} 
\end{figure*}

\begin{figure*}[h]
\centering
\includegraphics[scale = 0.52]{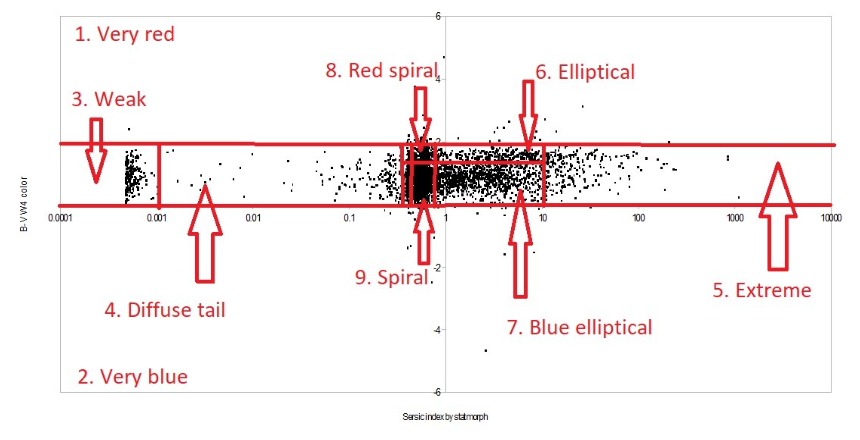}
\caption{Non-fake regions for bimodality test.}
\label{fig:bm2}    
\end{figure*}

\begin{figure*}[h]
\centering
\includegraphics[scale = 0.6]{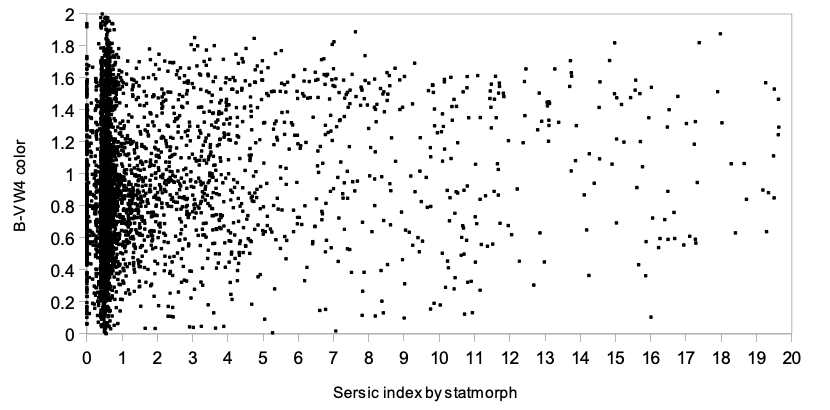}
\caption{Realistic region of color and Sersic index.}
\label{fig:bm3}   
\end{figure*}

\begin{table}
\begin{tabular}{| c | c | c | c | c | c | c |}
\hline
Region №	&	Region name	&	B-V min	&	B-V max	&	Sersic index min	&	Sersic index max	&	Number of galaxies	\\	
\hline
1	&	Very red	&	2	&	6	&	0	&	1095	&	22	\\	
\hline
2	&	Very blue	&	-6	&	0	&	0	&	1095	&	40	\\	
\hline
3	&	Weak	&	0	&	2	&	0.0004	&	0.001	&	184	\\	
\hline
4	&	Diffuse tail	&	0	&	2	&	0.001	&	0.35	&	169	\\	\hline
5	&	Extreme	&	0	&	2	&	10	&	1095	&	284	\\	
\hline
6	&	Elliptical	&	1.3	&	2	&	0.7	&	10	&	356	\\	
\hline
7	&	Blue Elliptical	&	0	&	1.3	&	0.7	&	10	&	1132	\\	\hline
8	&	Red spiral	&	1.3	&	2	&	0.45	&	0.7	&	253	\\	
\hline
9	&	Spiral	&	0	&	1.3	&	0.45	&	0.7	&	1446	\\	
\hline
10	&	Fake elliptical	&	1.3	&	2	&	0.35	&	0.45	&	85	\\	\hline
11	&	Fake spiral	&	0	&	1.3	&	0.35	&	0.45	&	395	\\	\hline
\end{tabular}
\caption{Regions in the distribution of color index - Sersic index for VIPERs galaxies shown at Fig.\ref{fig:bm2}}
\end{table}

\hfill \break
Except of needed bimodality (density excess in regions 9 and 6 for elliptical and spiral galaxies), distribution at Fig. 1. is affected by two artifact from the algorithm of Sersic index calculation. The first is 'weak' region N3 at Fig. 5. It originate from the lower bound of Sersic index in statmorph. The second artifact, 'fake regions' N10 and N11 (Table 1 and Fig.5) appeared after our special modification of some images. The reason of such modification os the following. For 40\% of galaxies statmorph can not calculate Sersic index and set its value ns=1. To force statmorph to select another value of ns, two modifications of image described above were applied. Here we will describe it once more in more details.

\begin{figure*}[h]
\includegraphics[width=1\linewidth]{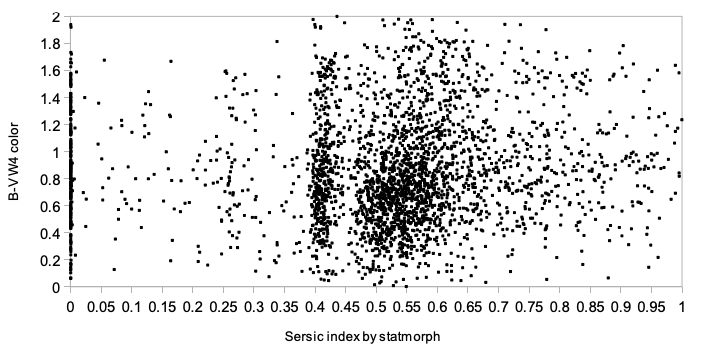}
\caption{Subregion for detection of fake values of Sersic index.}
\label{fig:bm4} 
\end{figure*}

\begin{figure*}[h]
\includegraphics[width=1\linewidth]{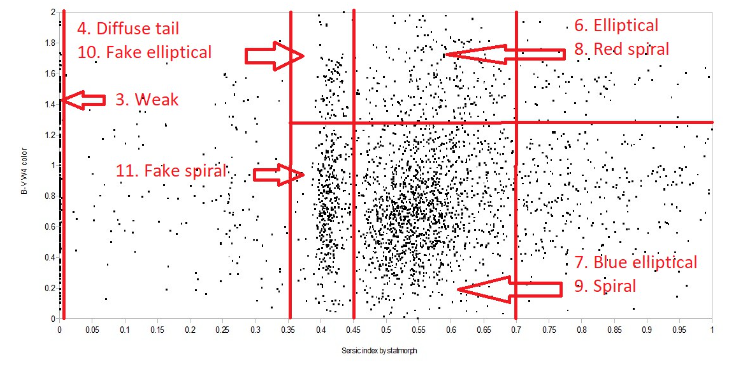}
\caption{Two regions (10 and 11) with incorrect calculation of Sersic index.}
\label{fig:bm5}    
\end{figure*}

\break

1. Segmentation map (aperture) was set as circle instead of calling recommended photutils python function. This gives realistic Sersic index for most problematic images.
\newline 2. If statmorph still cannot find Sersic index, round gaussian nucleus was added to image. After artifical deformation of galaxy image its Sersic index becomes $0.4\pm 0.02$.
Visual inspection of galaxies from all 11 regions did not allowed to explain fails of statmorph Sersic calculation. All regions include visually normal images, interacting galaxies, very faint images and small number of images with light pollution.
\newline Now with all these regions at Sersic-color distributions let's proceed to bimodality test. Overdensity in 'spiral' region N9 is obvious. Some of 85 galaxies in region 10 should have ns=1-4 so overdensity in region 6 will be underestimated. Nevertheless we suppose that excluding both regions 10 and 11 we will have still enough number of galaxies and correct proportion to detect the concentration of elliptical galaxies. The idea of test is the comparison of fraction of red galaxies for two ranges of ns: 0.45-0.7 and 0.7-10. These fractions are equal to 15\% and 24\% correspondingly. By this way we find that excess of elliptical galaxy number in region 6 is equal to 134 galaxies.

\newpage

\subsection{PSF test}

To check the influence of PSF model on Sersic index we considered Gaussian PSF model with different values of with parameter sigma. Results are shown at Figs. 10-13. X value is Sersic index calculated by GALFIT \cite{Krywult2017} and y value is Sersic index calculated by statmorph. We found that optimal PSF parameter is sigma=2 pixels. This value was used in all other Sersic index calculations by statmorph.

\begin{figure}[!h]
\centering
\begin{minipage}[t]{80 mm}
\centering
\includegraphics[scale = 0.45]{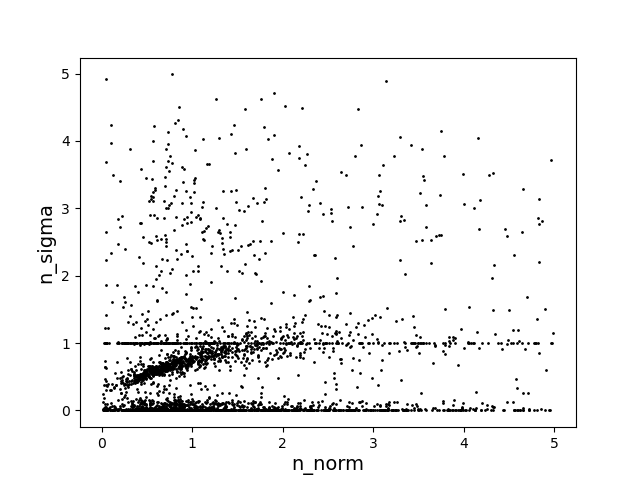}
\caption{Sigma=1.}\label{psf1}
\end{minipage}
\hfill
\begin{minipage}[t]{80 mm}
\centering
\includegraphics[scale = 0.45]{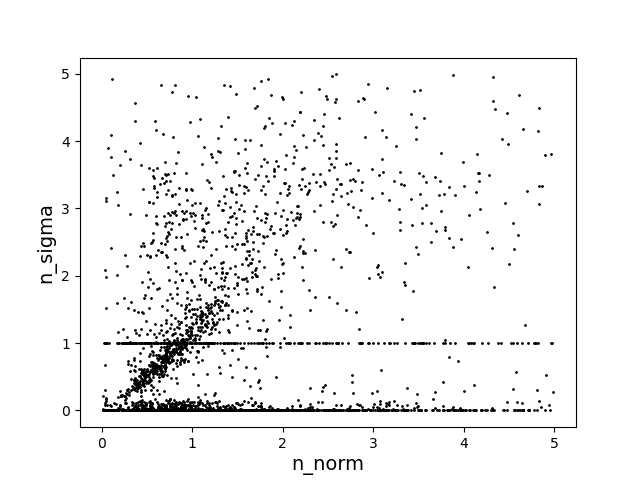}
\caption{Sigma=2.}\label{psf2}
\end{minipage}
\end{figure}

\begin{figure}[!h]
\centering
\begin{minipage}[t]{80 mm}
\centering
\includegraphics[scale = 0.45]{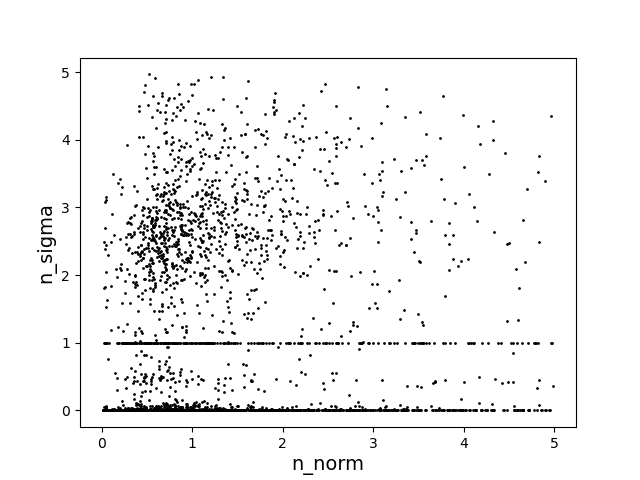}
\caption{Sigma=3.}\label{psf3}
\end{minipage}
\hfill
\begin{minipage}[t]{90 mm}
\centering
\includegraphics[scale = 0.45]{img/psf_test/psf_sigma_4.png}
\caption{Sigma=4.}\label{psf4}
\end{minipage}
\end{figure}

\newpage

\section{Results}

We calculated Gini and M20 statistics as the main advances morphological parameters.
We plot these parameters at for VIPERS sample (Fig. 14) and find out that galaxies can be divided to spiral, elliptical and merging. The same division was presented by \cite{Rodriguez-Gomez2019}.

This division is performed by the usage of so called buldge statistics and merger statistics.

Buldge statistics indicates morphological type of galaxy. According to \cite{Rodriguez-Gomez2019} it is calculated by the formula

$F = -0.693M_{20} + 4.95G - 3.96$

The value F=0 is the bound between spiral and elliptical galaxies. $F>0$ corresponds to elliptical galaxies and $F<0$ corresponds to spiral galaxies. To compare this division with the values of Sersic index we built Figs. 15 and 16. The value of Sersic index n=1 corresponds to spiral galaxies and n=3 corresponds to elliptical galaxies. We can see some correlation between n and F.

Merger statistics is calculated by the formula \cite{Rodriguez-Gomez2019}:

$S = 0.139M_{20} + 0.99G - 0.327$

$S>0$ corresponds to merging galaxies. We also built distribution of Merger statistics from Sersic index at Figs. 17 and 18. We can not find correlation of merger activity with Sersic index.

The differences between our results for VIPERS and Gini-M20 distribution for PanStarrs galaxies at $z<0.5$ could be explained it by cosmological evolution of galaxies. We found out that in modern Universe there are much more elliptical galaxies than at $z>0.5$ which corresponds to VIPERS sample. Also we concluded that galaxy mergers were more frequent in the early Universe.

\begin{figure*}[h]
\includegraphics[width=1.05\linewidth]{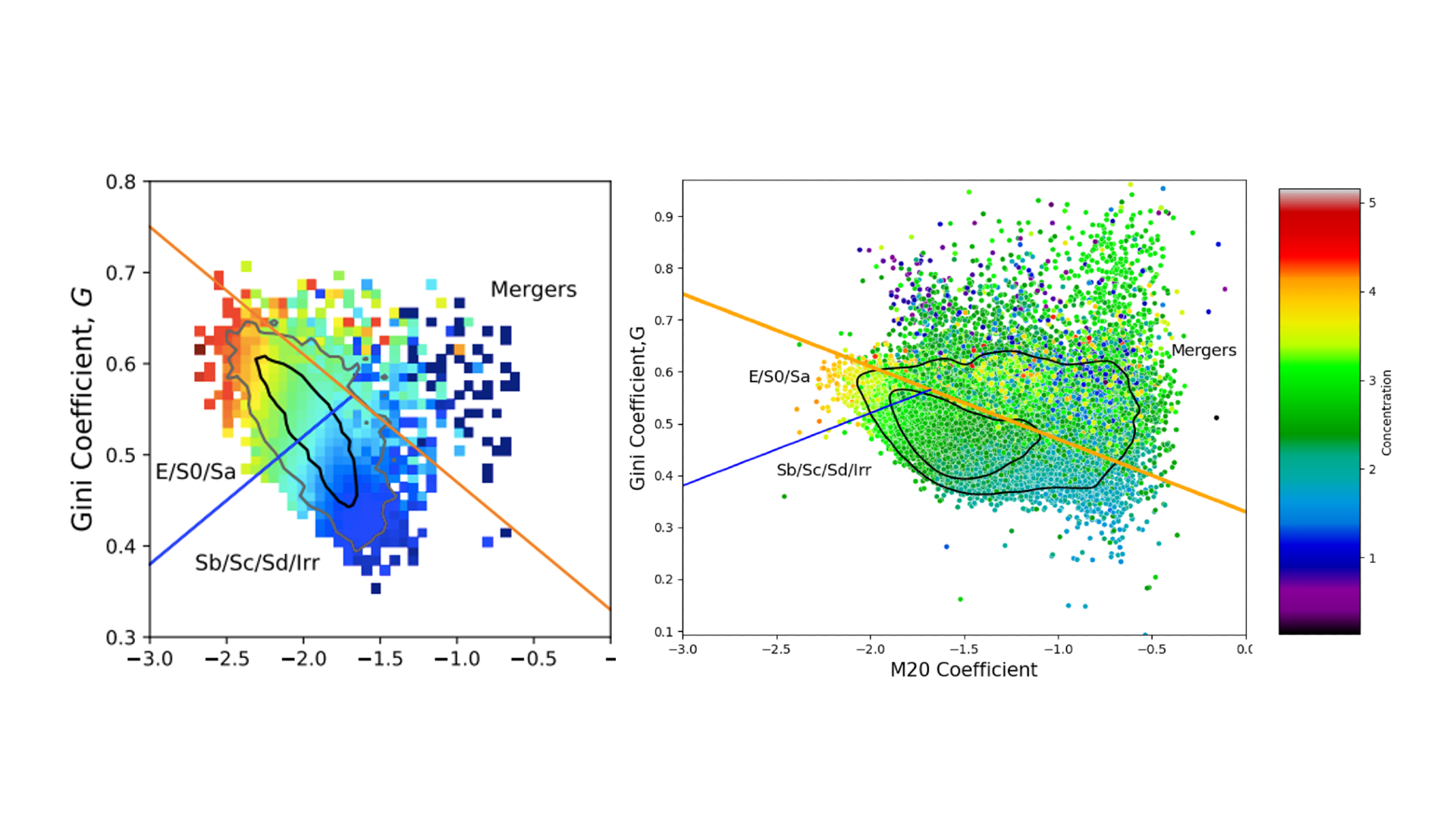}
\caption{Distribution of Gini and M20 parameters. Right panel - 70821 VIPERS galaxies. Left panel - simulated galaxies by \cite{Rodriguez-Gomez2019}. x axis - M20 parameter. y axis - Gini parameter. Blacker dots corresponds to higher values of concentration parameter C.}
\label{fig:bm5}    
\end{figure*}

\newpage

\subsection{Sersic index comparison}

To test our results we plotted buldge and merger statistics from Sersic index for test sample of 4388 galaxies. We considered both versions of Sersic index from our statmorph run and previous results from GALFIT. 

\begin{figure*}[h]
\centering
\includegraphics[scale = 0.49]{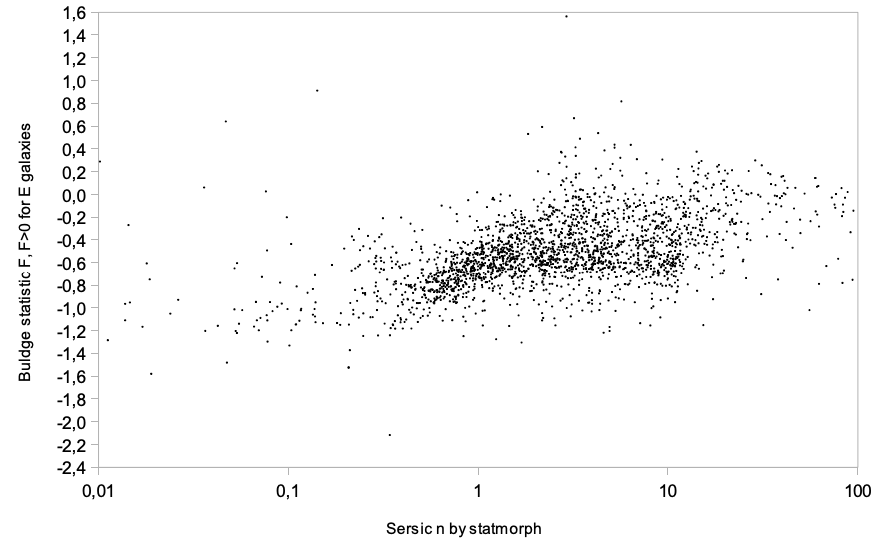}
\caption{Buldge statistic from Sersic index for test sample of VIPERS galaxies. Sersic index was calculated by statmorph.}
\label{fig:15}
\end{figure*}

\newpage

\begin{figure*}[h]
\centering
\includegraphics[scale = 0.49]{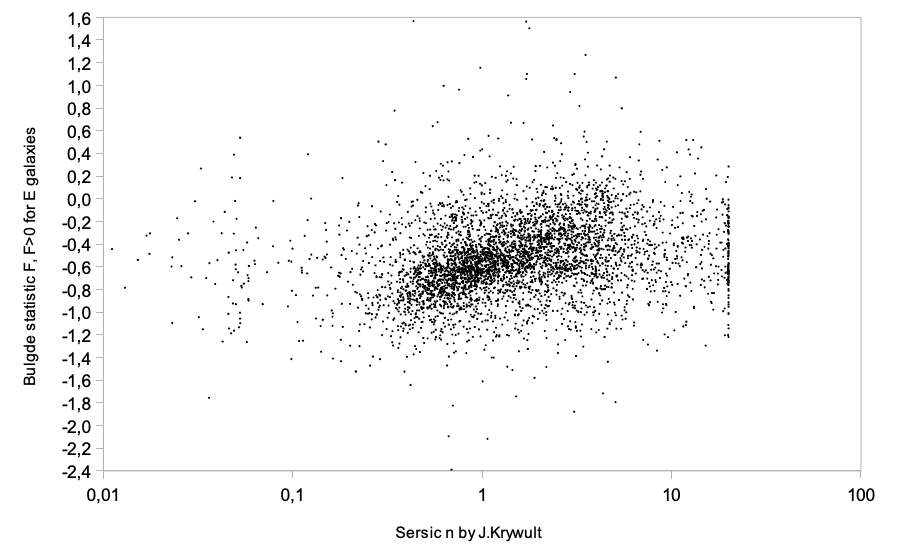}
\caption{Buldge statistic from Sersic index for test sample of VIPERS galaxies. Sersic index was calculated by GALFIT \cite{Krywult2017}.}
\label{fig:16}
\end{figure*}

\begin{figure*}[h]
\centering
\includegraphics[scale = 0.39]{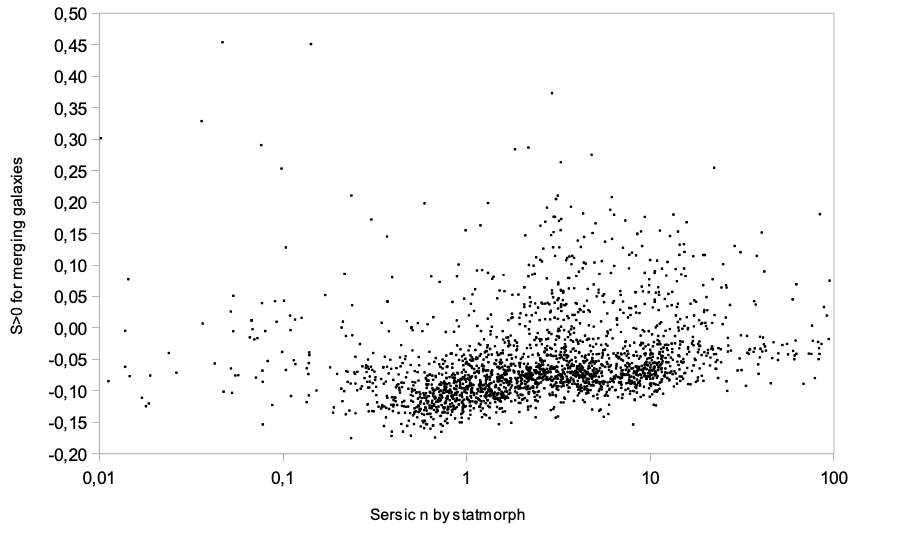}
\caption{Merger statistic from Sersic index for test sample of VIPERS galaxies. Sersic index was calculated by statmorph.}
\label{fig:17}
\end{figure*}

\newpage

\begin{figure*}[h]
\centering
\includegraphics[scale = 0.49]{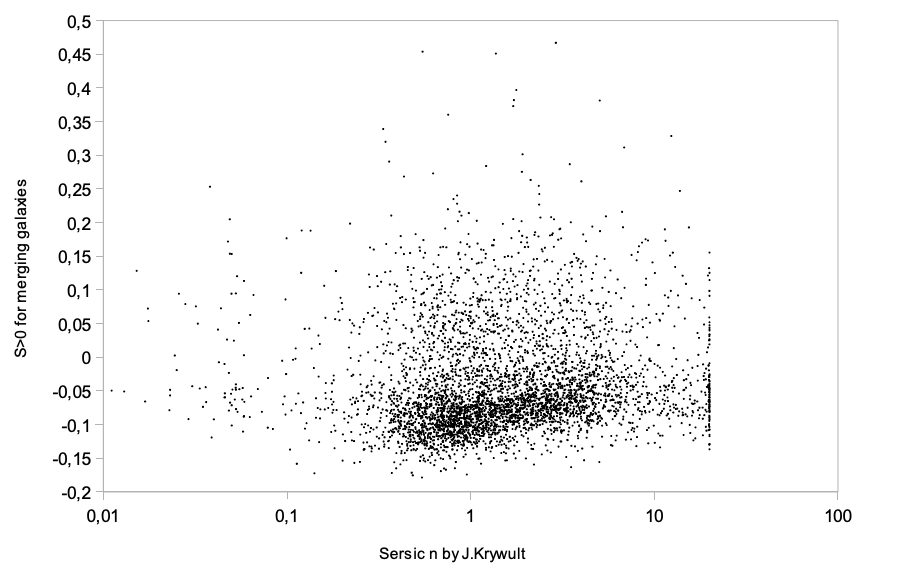}
\caption{Merger statistic from Sersic index for test sample of VIPERS galaxies. Sersic index was calculated by GALFIT \cite{Krywult2017}.}
\label{fig:18}   
\end{figure*}

\newpage

\subsection{Error analysis}

To evaluate errors of morphological parameters we simulated a number of different galaxy images with random background.
	Sersic power index was set as n=1.5 for all simulations. Four effective radii r were considered - 3 pixels, 6 pixels, 12 pixels and 24 pixels that corresponds 0.6, 1.1, 2.2 and 4.5 arcsec. Galaxy apparent magnitude was simulated by the amplitude a of Sersic model by the following way. The relation between the mentioned values is i=20-2.5lg(a/50). Foe example, a=1 for m=24.5; a=3 for m=23; a=9 for m=21.5; a=27 for m=20; a=81 for m=17.5 and a=243 for m=17.
	Size of poststamp with image is 100 pixels = 18.57 arcsec.
	
	Procedure of calculating of random errors was the following. For the given Sersic parameters a and r, random noise was generated 90 times. Mean value and standard deviation $\sigma $ was calculated for each morphological parameter. Exponential trends for all deviations are presented at Table 1. In each fit we used 11 values of magnitude from i=19 to i=24. These trends may be used as errors of morphological parameters for different magnitudes and radii. 
	
	Notes for evaluation the errors.
	
	1. Total number of mock images for each r is 11*90-990 round galaxies for Gini, M20, C and S and additional 990 images for asymmetry calculations. Statmorph often can not find parameters for weak galaxies with r=3 pixels. For r=6, 12 and 24 pixels we generated 3*2*990=5940 mock galaxies. To justify a number of mock objects, all parameters were calculated 100.000 times for r=6 and i=20 (Fig. 22).
	
	2. Since we run statmorph at images with no redshift needed, z is not a parameter for sample selection. Current results allows to expect major problems with $i>24$ and $r<1$ arcsec. Plots of errors $\sigma (i)$ are presented at Fig. 19-21 for reliability estimation.
	
	3. Blending effect can be estimated for asymmetric images by changing the distance between two components.

\begin{figure}[!h]
\centering
\begin{minipage}[t]{60 mm}
\centering
\includegraphics[scale = 3]{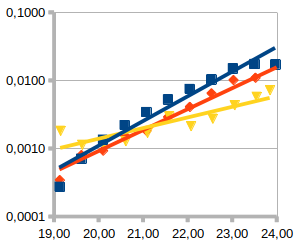}
\label{}
\end{minipage}
\hfill
\begin{minipage}[t]{100 mm}
\centering
\includegraphics[scale = 3]{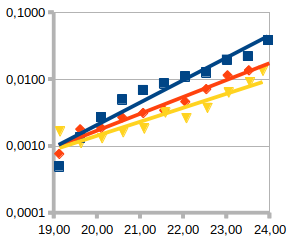}
\label{}
\end{minipage}
\caption{Errors of Gini and M20 parameters. Galaxy radius is 6 pixels for blue fit, 12 pixels for orange fit and 24 pixels for yellow fit.}
\end{figure}

\newpage

\begin{figure}[!h]
\centering
\begin{minipage}[t]{60 mm}
\centering
\includegraphics[scale = 2.5]{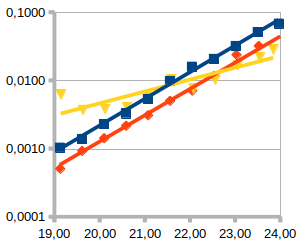}
\label{}
\end{minipage}
\hfill
\begin{minipage}[t]{100 mm}
\centering
\includegraphics[scale = 2.5]{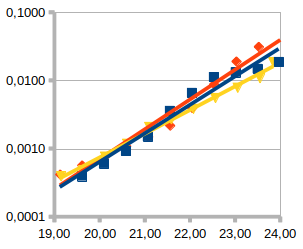}
\label{}
\end{minipage}
\caption{Concentration and smoothness errors as a function from i magnitude.}
\end{figure}

\begin{figure*}[h]
\centering
\includegraphics[scale = 2.5]{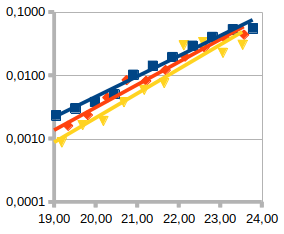}
\caption{Asymmetry errors.}
\label{}   
\end{figure*}

\begin{figure*}[h]
\centering
\includegraphics[scale = 0.3]{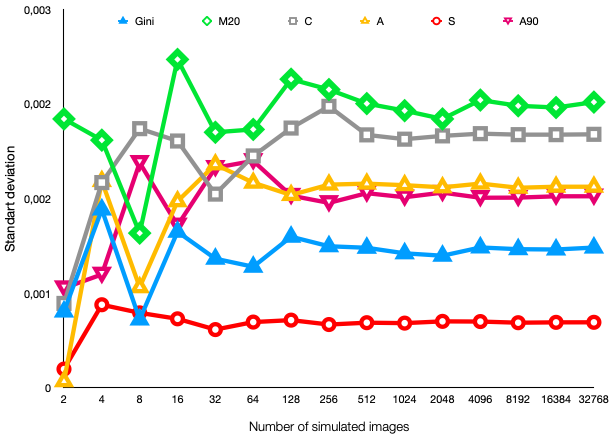}
\caption{Asymptotic limit of errors with increasing number of simulations. Errors of Gini are marked by blue triangles, M20 — green diamonds, Concentration — grey squares, Asymmetry — yellow triangles, Smoothness - red circles, asymmetry rotated by 90 degrees - pink triangles.}
\label{}   
\end{figure*}

\newpage

\subsection{Merger simulation}
As it was mentioned in previous sections, morphological parameters of galaxies are widely used in astrophysics. For example, it can be used for analysing galaxy mergers and interactions. We performed simulations of pairs of images without taking into account physical interaction of galaxies and changes of their shape. On Figures 23-25 one can see variations of statmorph parameters for mock images of galaxy pairs. Offset (distance) between centers of galaxies is measured in pixels, 19 pixels = 3.6 arcsec. Error bars shows standard deviation for 50 realisations of background. Brightness of central galaxy corresponds to i=22 at VIPERS images. Figures 23-25 demonstrates changes of values and errors of morphology parameters for close pairs of images. On Fig.26-27 are introduced images, which were built to calculate errors of asymmetry.
On Fig. 23 one may see variation of Gini and $M_{20}$ parameters. Change of Gini is the most inconspicuous, while $M_{20}$ is changing a lot during the merge, although resulting parameter remains constant. 
On Fig. 24 one may see variation of Smoothness and Asymmetry parameters. As in previous pictures, initial and final values (before and after merge) remains constant, but the change of them is different. Smoothness has something like maximum near 10 pixels offset(peak from 13 to 6). It can be explained by dividing of starting segmentation map by two maps of two different galaxies. Asymmetry, at the same time, changes weakly.
On the Figure 25 one may see variation of concentration parameter. Before merging(on 20 pixels distance), there was only one segmentation map, so the concentration was high. After decreasing the distance, segmentation map was divided into two regions, which higly decreased resulting concentration. But after merging, segmentation map had connected again, and concentration had returned to its maximum.
Errors on all of Figures were increasing during division of starting segmentation map on mid distances, and were decreasing during connection of two maps to one after merging on lower distances. All of mentioned processes can be seen on Figures 26-27, which shows simulation of merging of two galaxies.

\begin{figure}[!h]
\centering
\begin{minipage}[t]{60 mm}
\centering
\includegraphics[scale = 0.25]{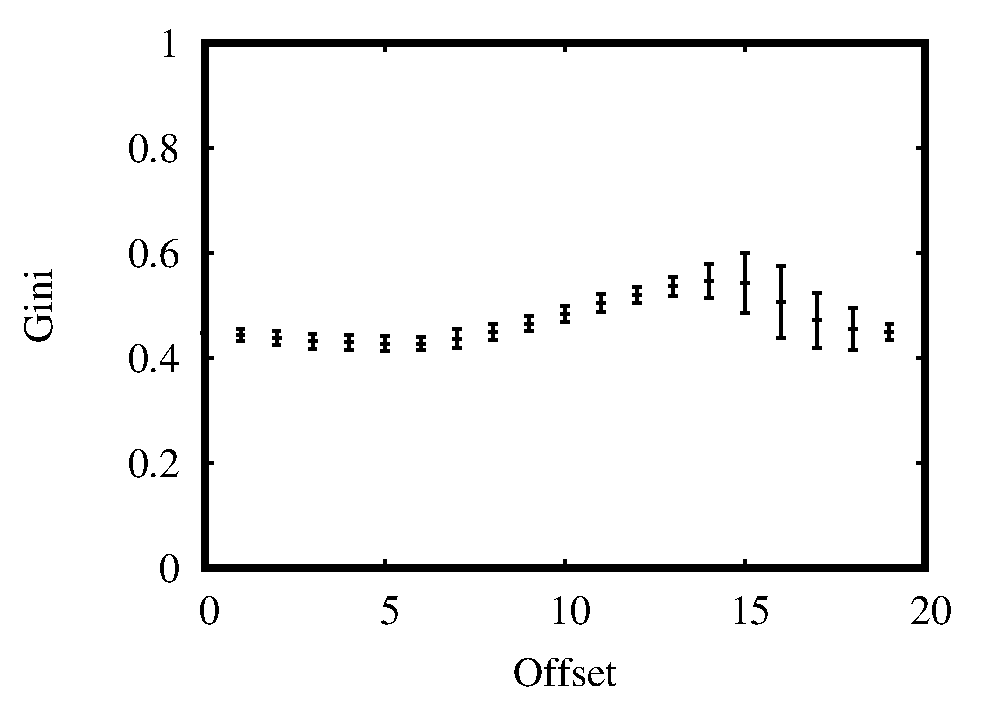}
\label{}
\end{minipage}
\hfill
\begin{minipage}[t]{100 mm}
\centering
\includegraphics[scale = 0.25]{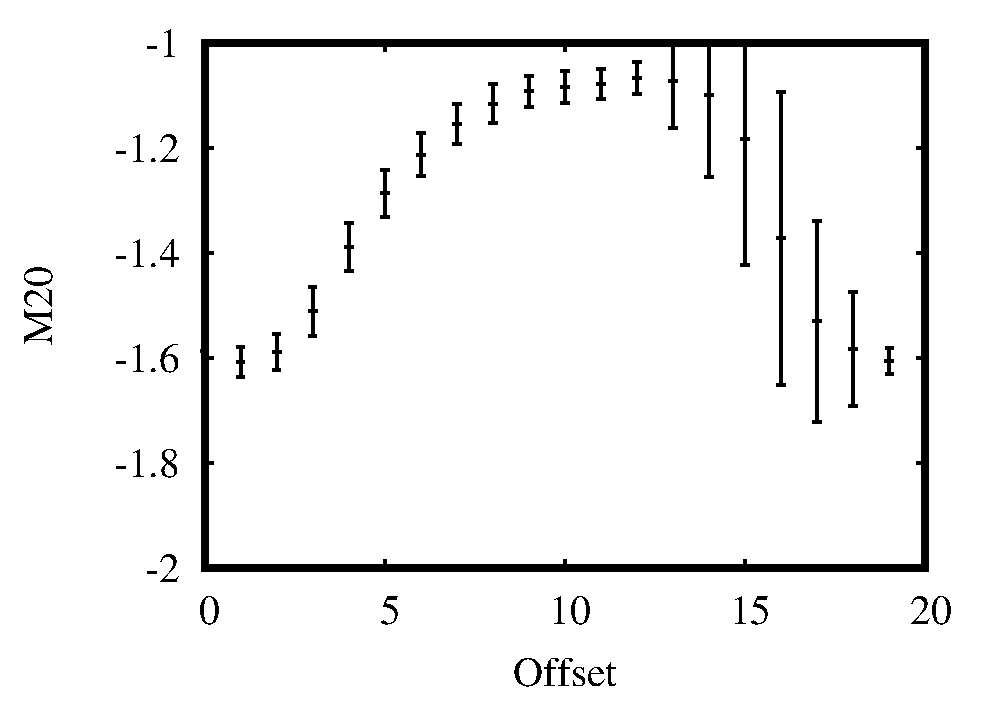}
\label{}
\end{minipage}
\caption{Variation of Gini and $M_{20}$ parameters of galaxy pairs with distance from their centers}
\end{figure}

\begin{figure}[!h]
\centering
\begin{minipage}[t]{60 mm}
\centering
\includegraphics[scale = 0.25]{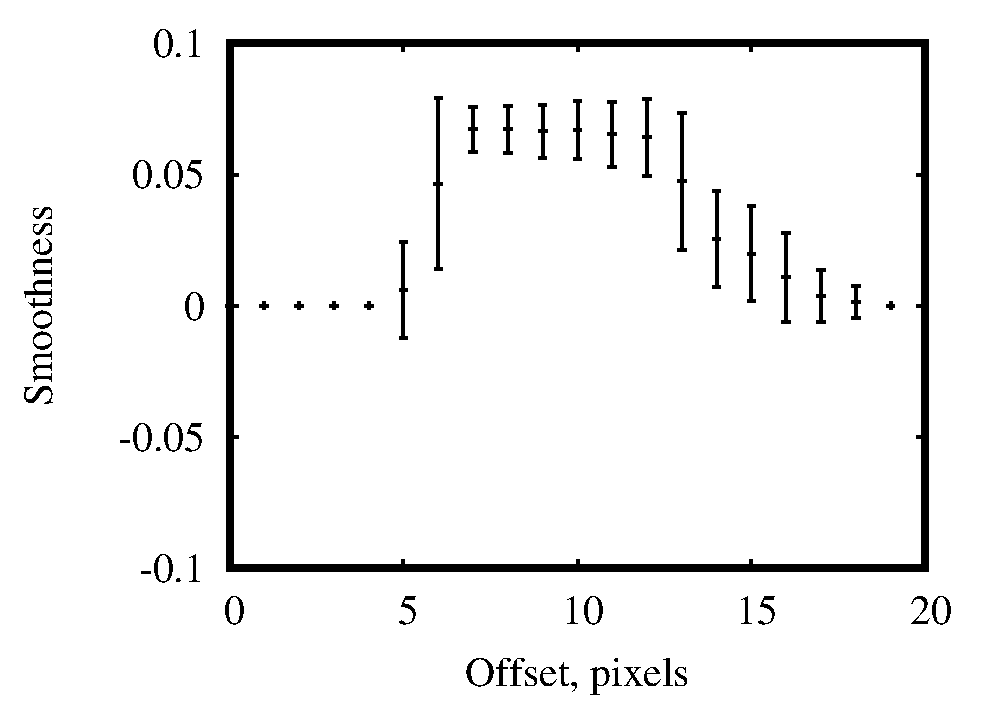}
\label{}
\end{minipage}
\hfill
\begin{minipage}[t]{100 mm}
\centering
\includegraphics[scale = 0.25]{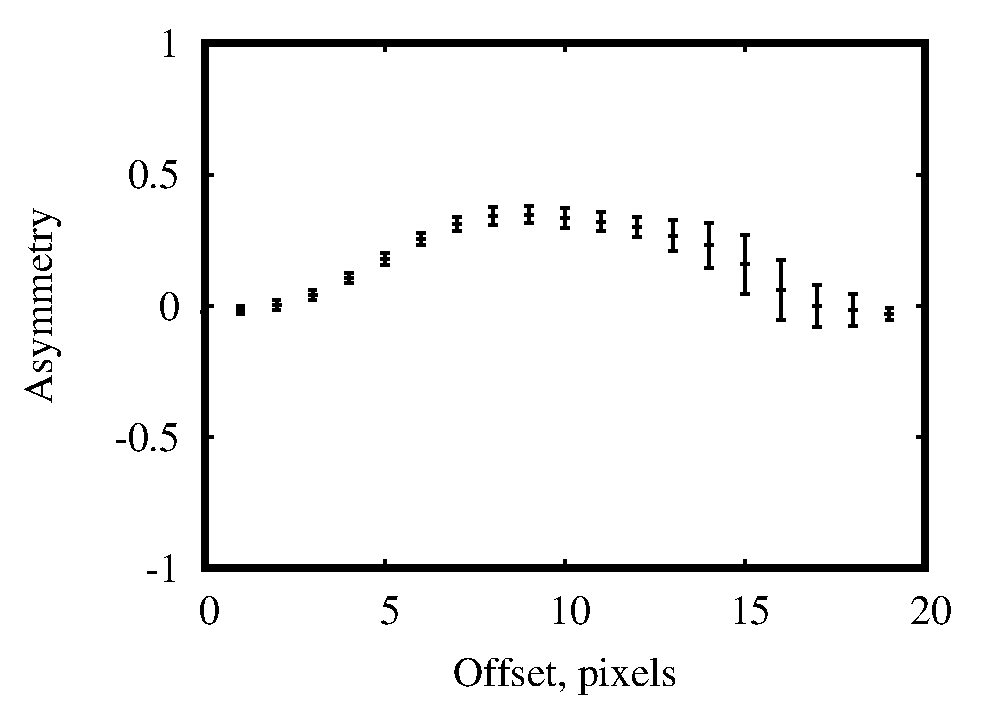}
\label{}
\end{minipage}
\caption{Variation of Smoothness and Asymmetry parameters of galaxy pairs with distance from their centers}
\end{figure}

\newpage

\begin{figure*}[h]
\centering
\includegraphics[scale = 0.25]{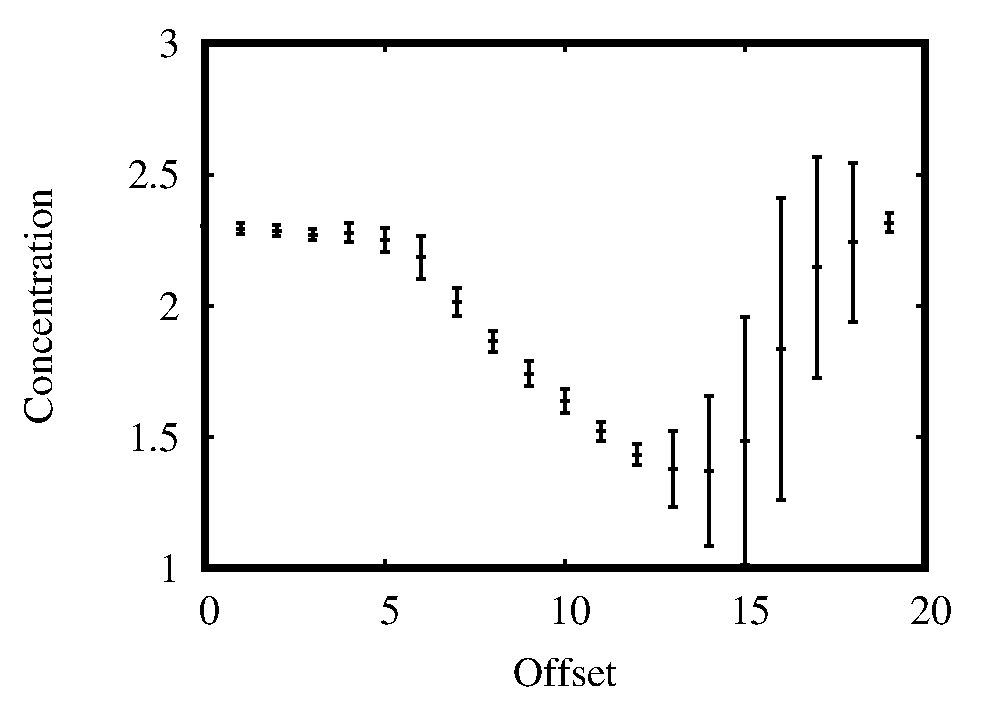}
\caption{Variation of Concentration of galaxy pairs with distance from their centers}
\label{}
\end{figure*}

\begin{figure}[!h]
\centering
\begin{minipage}[t]{30 mm}
\centering
\includegraphics[scale = 0.2]{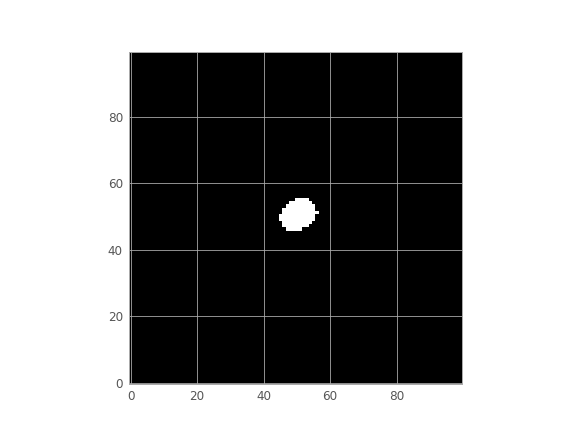}
\label{}
\end{minipage}
\hfill
\begin{minipage}[t]{30 mm}
\centering
\includegraphics[scale = 0.2]{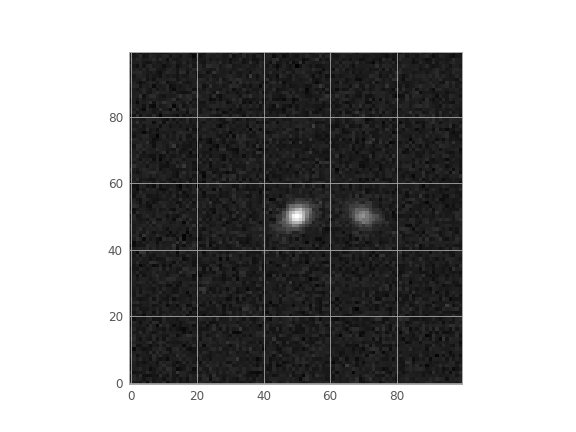}
\label{}
\end{minipage}
\hfill
\begin{minipage}[t]{30 mm}
\centering
\includegraphics[scale = 0.2]{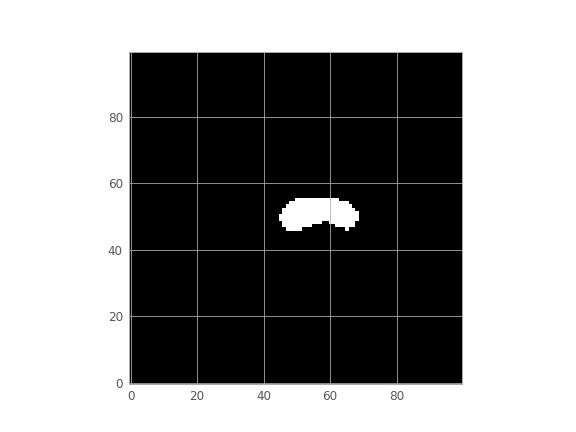}
\label{}
\end{minipage}
\hfill
\begin{minipage}[t]{30 mm}
\centering
\includegraphics[scale = 0.2]{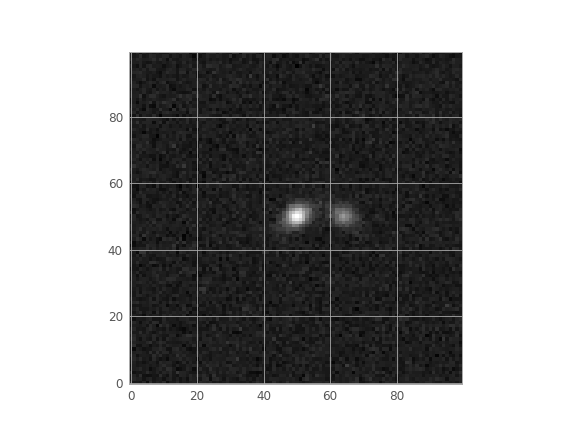}
\label{}
\end{minipage}
\caption{Segmentation maps and simulated images of close galaxies at distance of 19 pixels (left) and 14 pixels (right).}
\end{figure}

\newpage

\begin{figure}[!h]
\centering
\begin{minipage}[t]{30 mm}
\centering
\includegraphics[scale = 0.2]{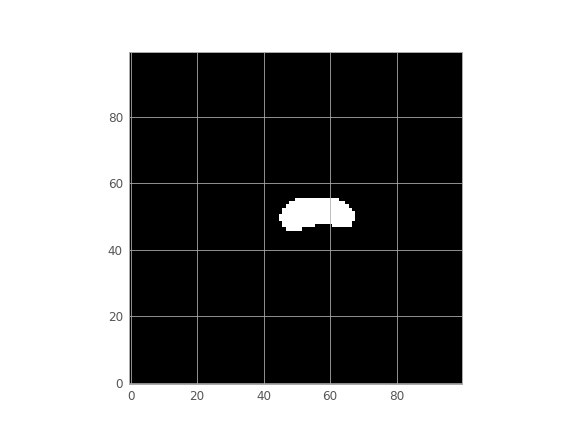}
\label{}
\end{minipage}
\hfill
\begin{minipage}[t]{30 mm}
\centering
\includegraphics[scale = 0.2]{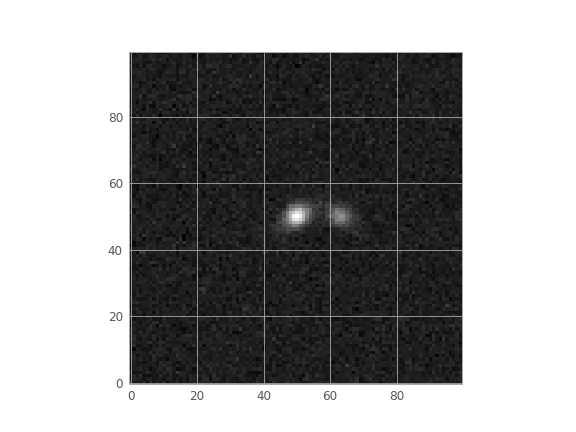}
\label{}
\end{minipage}
\hfill
\begin{minipage}[t]{30 mm}
\centering
\includegraphics[scale = 0.2]{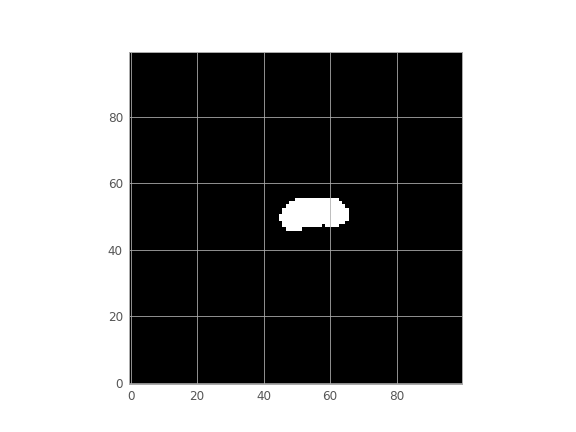}
\label{}
\end{minipage}
\hfill
\begin{minipage}[t]{30 mm}
\centering
\includegraphics[scale = 0.2]{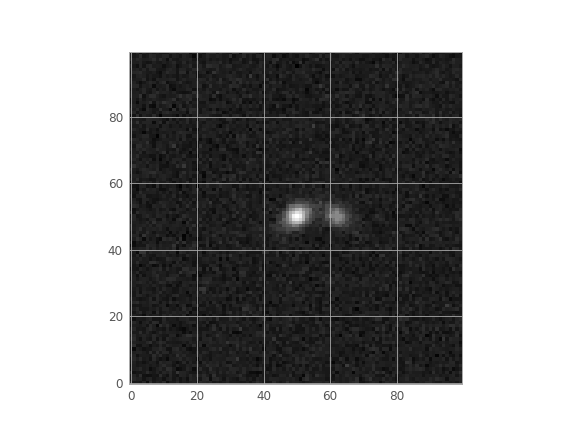}
\label{}
\end{minipage}
\end{figure}

\begin{figure}[!h]
\centering
\begin{minipage}[t]{30 mm}
\centering
\includegraphics[scale = 0.2]{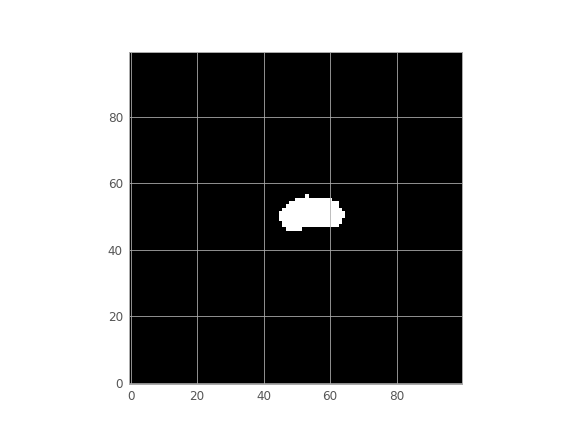}
\label{}
\end{minipage}
\hfill
\begin{minipage}[t]{30 mm}
\centering
\includegraphics[scale = 0.2]{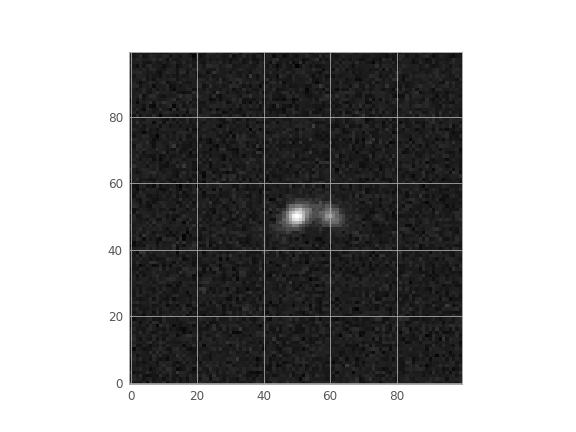}
\label{}
\end{minipage}
\hfill
\begin{minipage}[t]{30 mm}
\centering
\includegraphics[scale = 0.2]{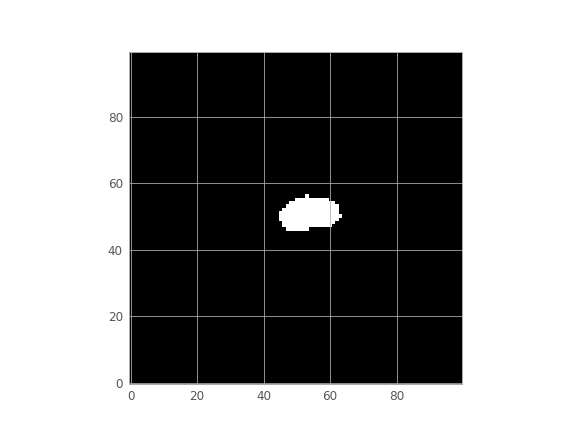}
\label{}
\end{minipage}
\hfill
\begin{minipage}[t]{30 mm}
\centering
\includegraphics[scale = 0.2]{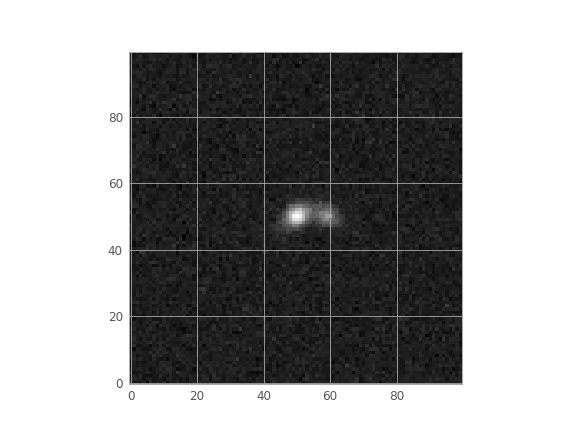}
\label{}
\end{minipage}
\end{figure}

\begin{figure}[!h]
\centering
\begin{minipage}[t]{30 mm}
\centering
\includegraphics[scale = 0.2]{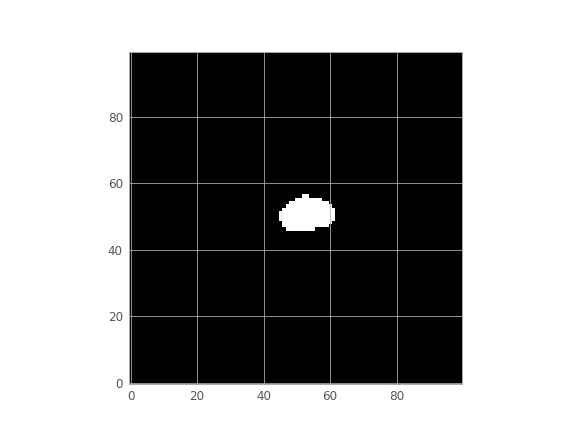}
\label{}
\end{minipage}
\hfill
\begin{minipage}[t]{30 mm}
\centering
\includegraphics[scale = 0.2]{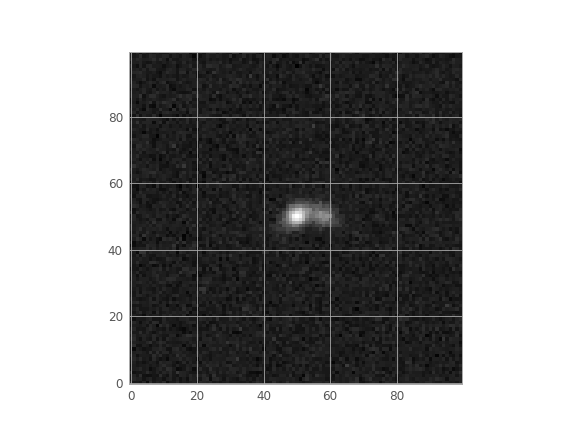}
\label{}
\end{minipage}
\hfill
\begin{minipage}[t]{30 mm}
\centering
\includegraphics[scale = 0.2]{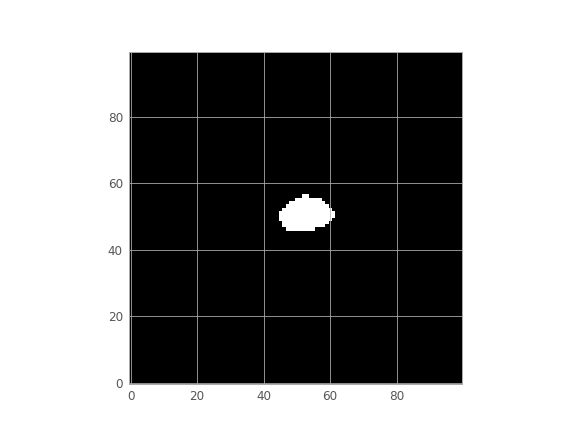}
\label{}
\end{minipage}
\hfill
\begin{minipage}[t]{30 mm}
\centering
\includegraphics[scale = 0.2]{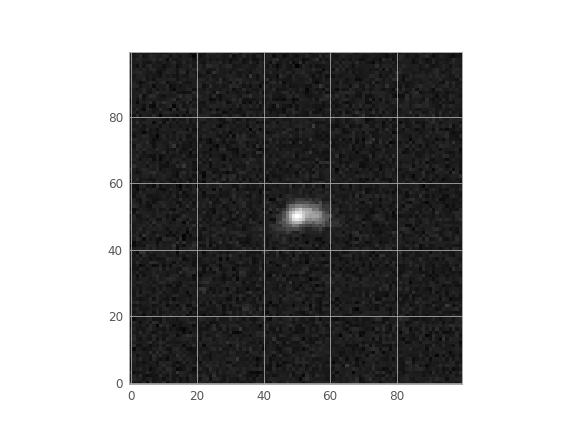}
\label{}
\end{minipage}
\end{figure}

\begin{figure}[!h]
\begin{minipage}[t]{30 mm}
\centering
\includegraphics[scale = 0.2]{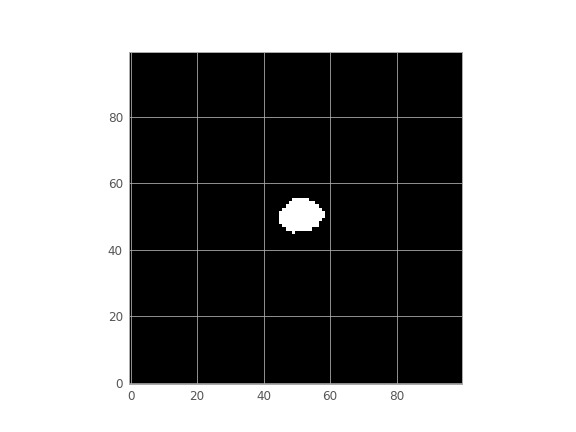}
\label{}
\end{minipage}
\hfill
\begin{minipage}[t]{30 mm}
\centering
\includegraphics[scale = 0.2]{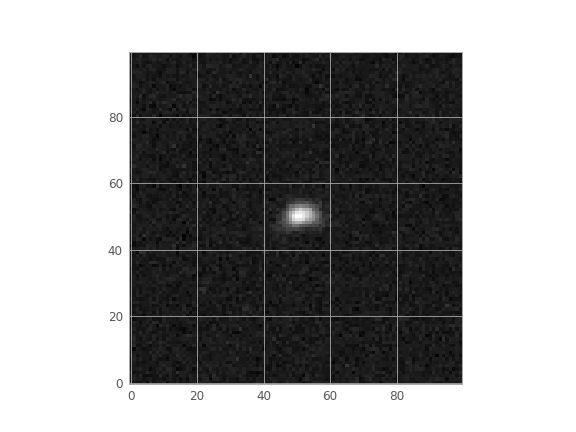}
\label{}
\end{minipage}
\hfill
\begin{minipage}[t]{30 mm}
\centering
\includegraphics[scale = 0.2]{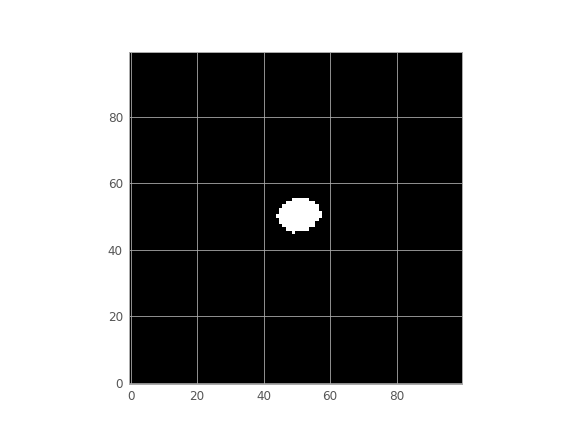}
\label{}
\end{minipage}
\hfill
\begin{minipage}[t]{30 mm}
\centering
\includegraphics[scale = 0.2]{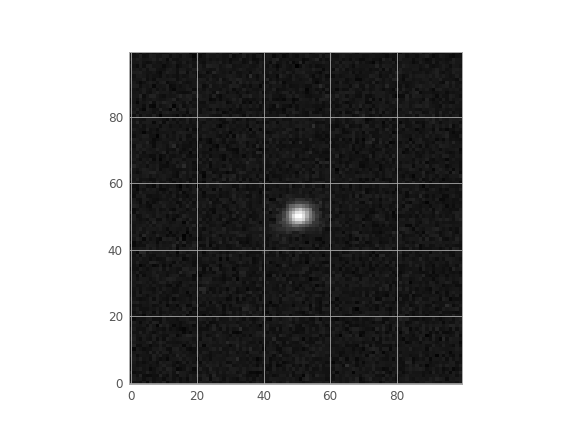}
\label{}
\end{minipage}
\caption{Simulation of segmentation maps and images of close galaxies. Distance between galaxy centers is equal to 13, 11, 9 and 5 pixels for the left column. Distance between galaxy centers is equal to 12, 10, 8 and 3 pixels for the right column.}
\end{figure}

\newpage

\section{Discussion}


We have to compare our results not only with similar results by \cite{Rodriguez-Gomez2019} for simulated galaxies, but also with many more papers in which these parameters were obtained for different samples of galaxies. 

1. Gini-M20 classification is widely used in many papers about galaxy morphology, p.e. in Lotz, Primack \& Madau, arxiv/0311352 (LPM04)(Fig. 26.).On this figure they have built Gini-M20 statistics and explained all elements on it: red circles:E/S0, green triangles:Sa-Sbc, blue crosses:Sc-Sd, diamonds:dI, bars:edge-on spirals). As one can see, almost all galaxies lie below the dashed line.
\begin{figure*}[h]
\includegraphics[width=1\linewidth]{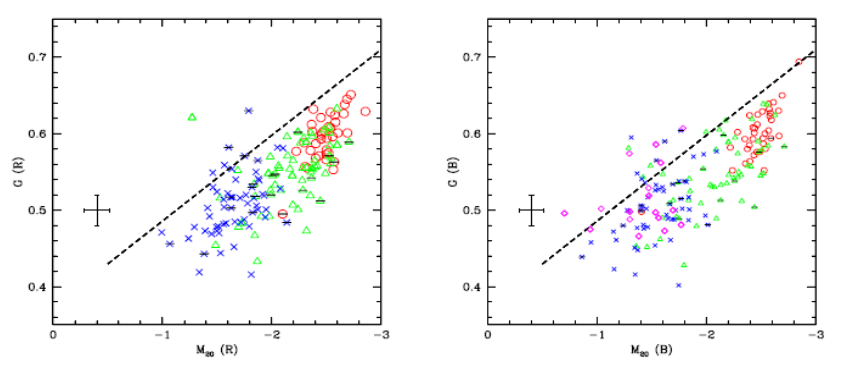}
\caption{Gini-M20 classification was introduced in Lotz, Primack \& Madau, 2004 \cite{Lotz2004}.}
\label{fig:dsc1} 
\end{figure*}

2.It was applied in \cite{Lotz2008} to galaxies, observed in All-wavelength Extended Groth Strip InternationalSurvey, AEGIS. (Hubble Telescope) also to find local merger candidates and to differ early and late-type galaxies.  $0.2<z<1.2$. Upper line(dotted green) also mentiones division between merger candidates from normal Hubble types, as we have on our resulting graph. Dotted green line divides early types and late types as well. On the right picture one can see evidence for bimodality between early and late types, as they are two zones of maximum in contours. Fig. 27-28.

\newpage

\begin{figure*}[h]
\includegraphics[width=1\linewidth]{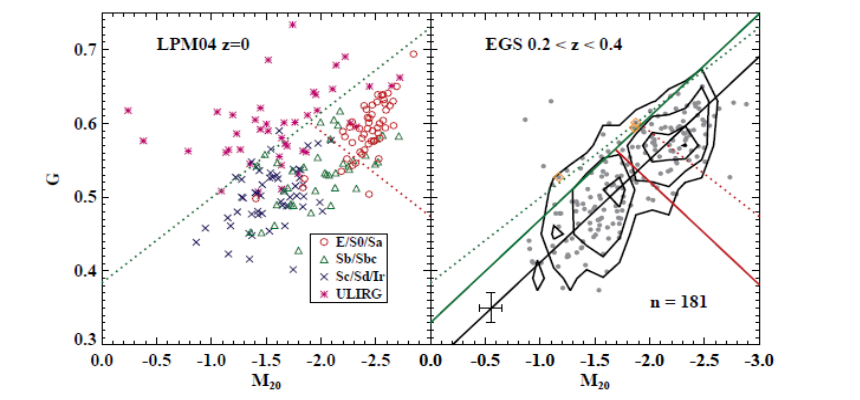}
\caption{Gini-$M_{20}$ classification, was introduced in Lotz et al.,2008}
   
\end{figure*}

\begin{figure*}[h]
\includegraphics[width=0.85\linewidth]{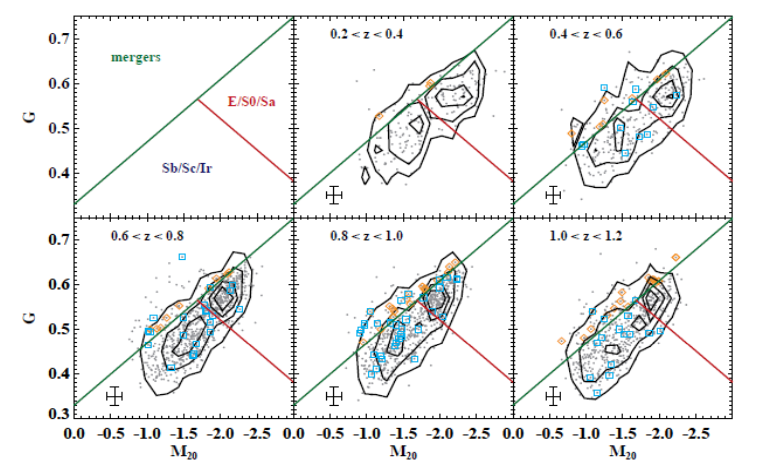}
\caption{Gini-$M_{20}$ classification with redshift dependence, was introduced in Lotz et al.,2008 }
\label{fig:dsc2} 
\end{figure*}

\newpage

3. Also this classification was used for merger diagnostics of simulated galaxies in \cite{Snyder2015a}. Fig. 29. On this pictures one can see dependence of morphological parameters G,$M_{20}$ and C with time, their in-time evolution. The main result, wich was suggested in paper, is that morphological evolution is not uniform. Darker contours enclose regions of increasing logarithm of relative number density, evenly spaced. Dotted lines are showing mergers/early/late-type division, as it was described above. 
\begin{figure*}[h]
\includegraphics[width=1\linewidth]{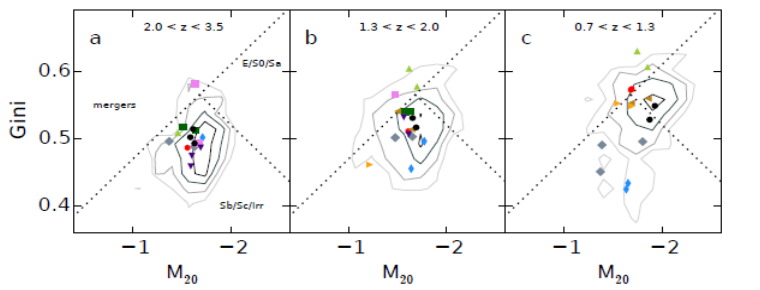}
\caption{Also this classification was used for merger diagnostics of simulated galaxies in \cite{Snyder2015a}. Fig. 29. }
\label{fig:dsc3}    
\end{figure*}

4. Buldge statistics F based on Gini-M20 classification was analyzed in (Snyder et al., 2015b) \cite{Snyder2015b} for Illustris simulation. Besides morphology classification, the following parameters were fitted with Buldge statistics: Star formation rate, stellar mass, galaxy size and galaxy rotation. Fig. 30-33.

\begin{figure*}[h]
\includegraphics[width=1\linewidth]{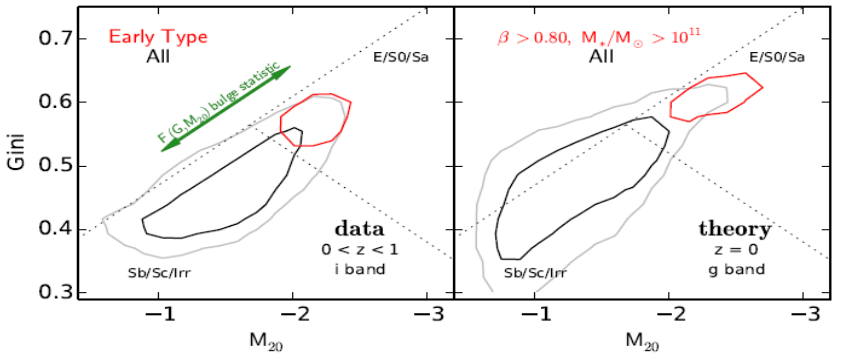}
\caption{Buldge statistics analysis on Gini-M20 distribution for Illustris simulation (Snyder et al., 2015b) \cite{Snyder2015b}. Correlation of buldge statistics with galaxy morphology type. }
\label{fig:dsc4}    
\end{figure*}

\begin{figure*}[h]
\includegraphics[width=0.4\linewidth]{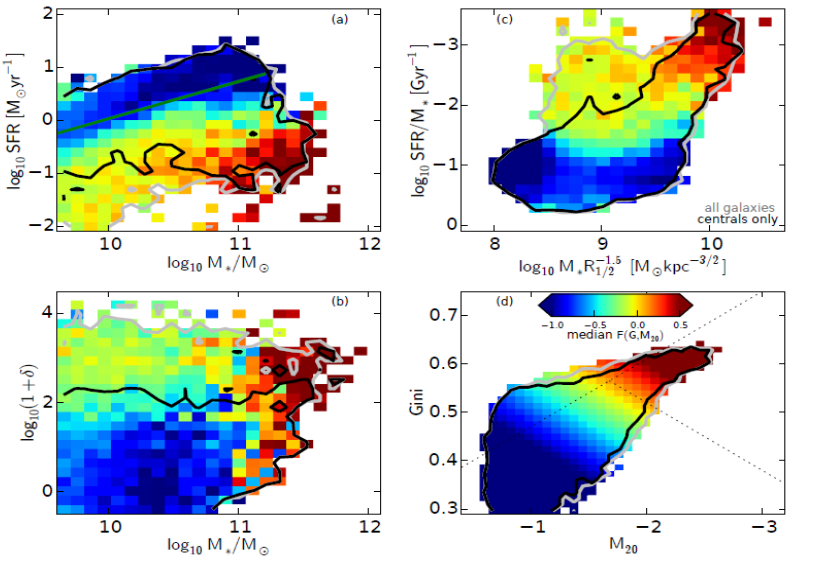}
\caption{Buldge statistics analysis on Gini-M20 distribution for Illustris simulation (Snyder et al., 2015b) \cite{Snyder2015b}. Correlation of buldge statistics with star formation rate and galaxy size. }
\label{fig:dsc5}    
\end{figure*}

\begin{figure*}[h]
\includegraphics[width=0.3\linewidth]{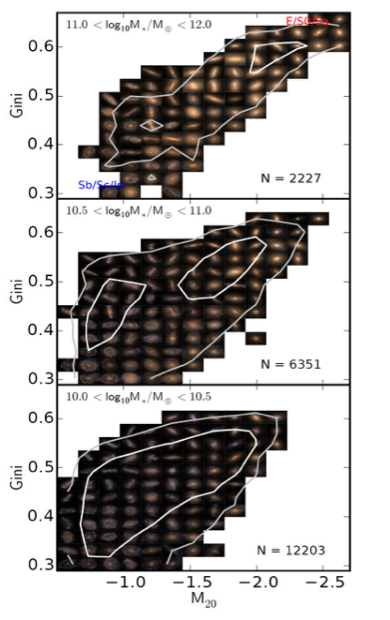}
\caption{Buldge statistics analysis on Gini-M20 distribution for Illustris simulation (Snyder et al., 2015b) \cite{Snyder2015b}. Correlation of buldge statistics with stellar mass. }
\label{fig:dsc6}    
\end{figure*}

\newpage

\begin{figure*}[h]
\includegraphics[width=0.5\linewidth]{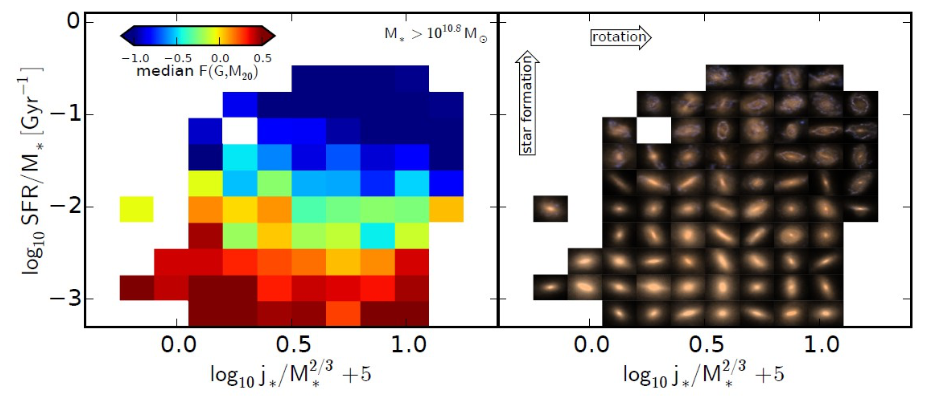}
\caption{Buldge statistics analysis on Gini-M20 distribution for Illustris simulation (Snyder et al., 2015b) \cite{Snyder2015b}. Correlation of buldge statistics with galaxy rotation. j* is angular momentum of galaxy. }
\end{figure*}

After comparison of our Gini and M20 parameters with many other works we confirm that our results are in agreement with results, obtained with another authors.

Secondly, we have got satisfied statistical errors for results, presented in our paper. Some disagreement (small differences) between our and other mentioned papers are within statistical errors ranges.

\indent Results, which are shown in this paper, can be used in further studies of influence of merging and environment on galaxy morphology, and for advanced classification of galaxies by their morphological parameters via statmorph code. Sersic index calculated by statmorph code in worse than the one by GALFIT [7], as it was shown in previous chapters. Statmorph Sersic index estimation is not suitable for galaxy morphology analysis.

\section{Acknowledgments}

Authors are thankfull to prof. Agniezka Pollo(Warsaw) for carefull scientific supervision of this study. This paper uses data from the VIMOS Public Extragalactic Redshift Survey (VIPERS). VIPERS has been performed using the ESO Very Large Telescope, under the "Large Programme" 182.A-0886. The participating institutions and funding agencies are listed at http://vipers.inaf.it . We are also thankfull to Dr. Janusz Krywult (Kielce) for valuable help with error analysis and testing of results.

\section{Appendix}

We present here at Figs. 36-38 illustrations of certain steps of galaxy image analysis by statmorph tool. Corresponding steps for simulated asymmetric image of merging galaxies are presented at Figs. 39-40. Corresponding images of real and simulated galaxies were used for analysis of morphological parameters and their errors (see Section 5.1.).

\begin{figure}[!h]
\centering
\begin{minipage}[t]{60 mm}
\centering
\includegraphics[scale = 0.4]{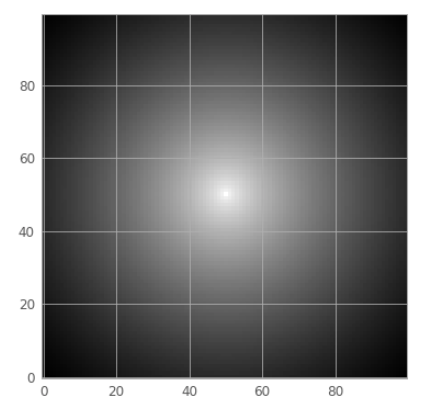}
\end{minipage}
\hfill
\begin{minipage}[t]{100 mm}
\centering
\includegraphics[scale = 0.4]{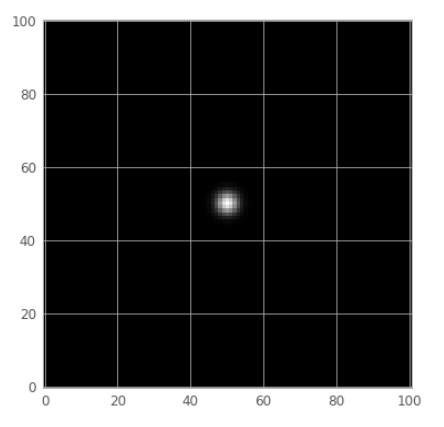}
\end{minipage}
\caption{On the left picture one may see round sersic profile. This image should be convolved with PSF. On the right one may see psf image}
\label{ap1}
\end{figure}

\begin{figure}[!h]
\centering
\begin{minipage}[t]{60 mm}
\centering
\includegraphics[scale = 0.4]{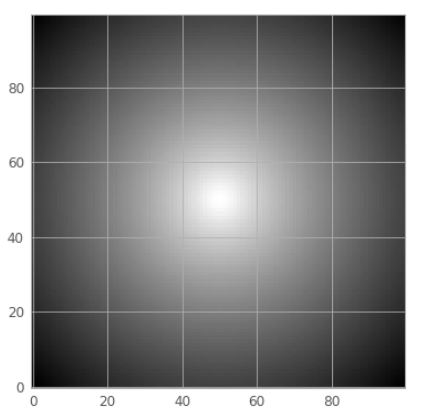}
\end{minipage}
\hfill
\begin{minipage}[t]{100 mm}
\centering
\includegraphics[scale = 0.4]{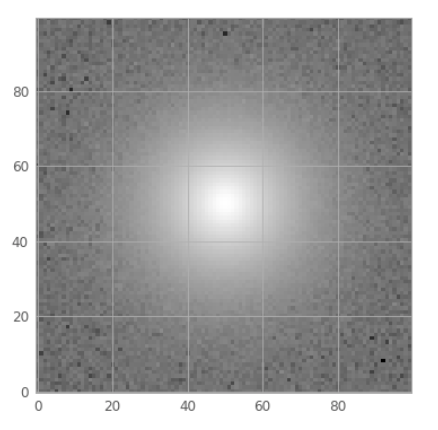}
\end{minipage}
\caption{These two pictures represent convolution of starting image with the psf and adding noise.}
\label{ap2}
\end{figure}

\begin{figure}[!h]
\centering
\begin{minipage}[t]{60 mm}
\centering
\includegraphics[scale = 0.4]{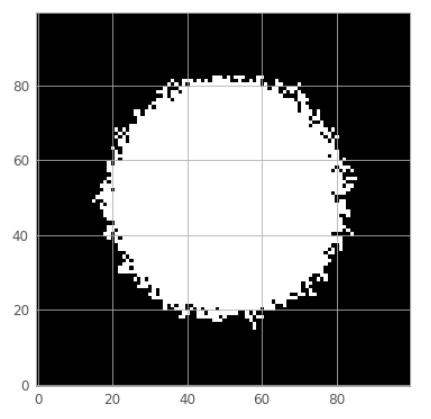}
\end{minipage}
\hfill
\begin{minipage}[t]{100 mm}
\centering
\includegraphics[scale = 0.4]{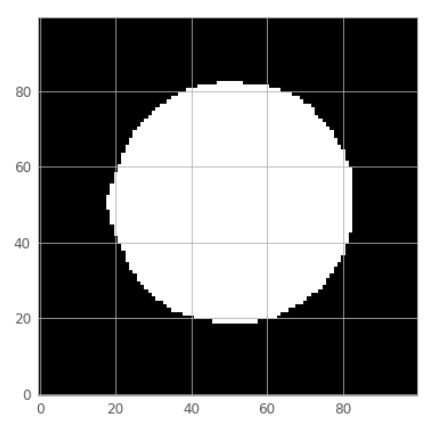}
\end{minipage}
\caption{These two pictures represent generation of the segmentation map and result of its smoothing.}
\label{ap3}
\end{figure}

\begin{figure}[!h]
\centering
\begin{minipage}[t]{60 mm}
\centering
\includegraphics[scale = 0.4]{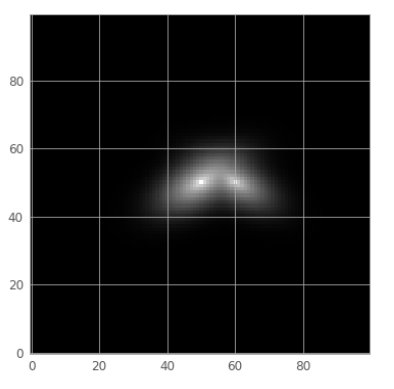}
\end{minipage}
\hfill
\begin{minipage}[t]{100 mm}
\centering
\includegraphics[scale = 0.4]{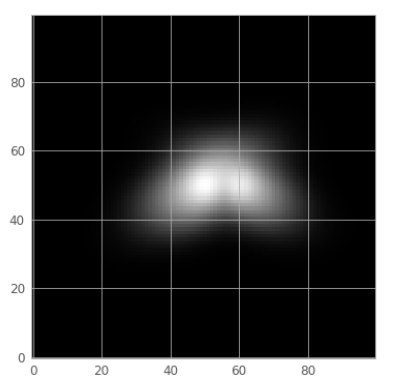}
\end{minipage}
\caption{These two pictures represent assymetric image and its convolution with the psf.}
\label{ap4}
\end{figure}

\begin{figure}[!h]
\centering
\begin{minipage}[t]{60 mm}
\centering
\includegraphics[scale = 0.4]{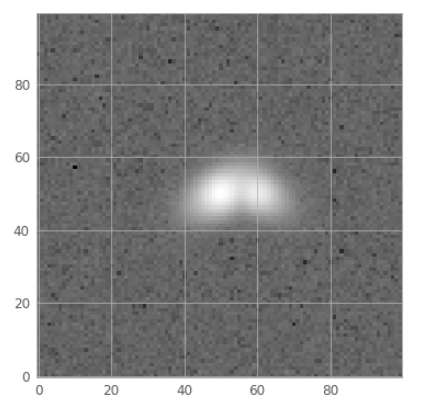}
\label{ap9}
\end{minipage}
\hfill
\begin{minipage}[t]{100 mm}
\centering
\includegraphics[scale = 0.4]{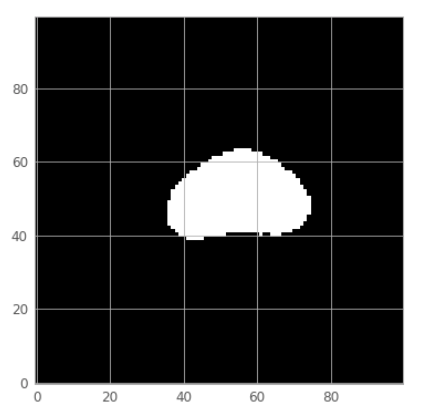}
\label{ap10}
\end{minipage}
\caption{These two pictures represent adding noise to assymetric image and generated segmentation map for it.}
\end{figure}

\end{document}